\documentclass[letterpaper,12pt]{extarticle}
\usepackage{graphicx}
\usepackage{dcolumn}
\usepackage{bm}
\usepackage{tabularx} 
\usepackage{preamble}
\usepackage{circledsteps}
\usepackage{breqn}
\usepackage{overpic}
\usepackage{circuitikz}
\usepackage[export]{adjustbox}
\usepackage{upgreek}
\usepackage{enumerate}
\usepackage{url}
\usepackage{bbm}
\usepackage{setspace}
\usepackage[font=small,labelfont=bf]{caption}
\usetikzlibrary{calc}
\usepackage{extarrows}
\usepackage{amsmath,amssymb}
\usepackage{booktabs,amsfonts,dcolumn}
\usepackage{circledsteps}
\usepackage{subcaption}
\usepackage{booktabs,subcaption,amsfonts,dcolumn}
\usetikzlibrary{decorations.pathmorphing}

\numberwithin{equation}{section}
\newcolumntype{d}[1]{D..{#1}}

\definecolor{refkey}{rgb}{0.9451,0.2706,0.4941}
\definecolor{labelkey}{rgb}{0.9451,0.2706,0.4941}
\usepackage{cite}
\usepackage{simplewick}

\def\z2{$\mathbb{Z}_2$}

\definecolor{darkgray}{rgb}{0.33, 0.33, 0.33}
\usepackage{braket}

\newcommand{\Tr}{{\mathrm{Tr}}}

\newcommand{\red}{}
\usepackage[left=30mm,right=30mm,top=30mm,bottom=30mm]{geometry}
\usetikzlibrary{arrows.meta,calc}
\usepackage{authblk}
\numberwithin{equation}{section}
\definecolor{MyGreen}{RGB}{10,120,10}
\usetikzlibrary{shapes.misc}
\tikzset{cross/.style={cross out, draw=black, minimum size=2*(#1-\pgflinewidth), inner sep=0pt, outer sep=0pt},
cross/.default={1pt}}
\definecolor{grayrow}{gray}{0.9}
\begin{document}

\onehalfspacing
\title{\Huge
Large-charge R\'enyi entropy
}

\author{Masataka Watanabe\footnote{\href{mailto:max.washton@gmail.com}{max.washton@gmail.com}}
}
\affil{\it \small The University of Tokyo, Tokyo 113-0033, Japan}
\date{}
\pagenumbering{gobble}

\maketitle

\thispagestyle{empty}
\begin{abstract}
\vspace{10mm}
\noindent
    The charged (symmetry-resolved) vacuum R\'enyi entanglement entropy on a disk is computed in the limit of large $U(1)$ global charge for any R\'enyi index.
We show that it behaves universally for a broad class of conformal field theories including the $O(2)$ Wilson-Fisher fixed-point, by using the effective field theory at large global charge. 
The result establishes one of the first concrete computations of entanglement quantities in strongly-coupled field theories.
\end{abstract}
\newpage
\pagenumbering{arabic}
\pagestyle{plain}
\onehalfspacing
\tableofcontents
\newpage
\setcounter{tocdepth}{2}
\maketitle
\pagenumbering{arabic}
\pagestyle{plain}

\section{Introduction}
\label{sec:intro}

Quantum information theory has greatly advanced our understanding of the quantum theory of gravity \it via \rm the holographic duality.
The Ryu–Takayanagi formula states that the entanglement entropy (EE) of a region is holographically dual to the area of the minimal surface in the bulk geometry which ends on the boundary of that region \cite{Ryu:2006bv}.
It suggests an idea that quantum information-theoretic quantities plays a fundamental role in the emergence of spacetime.

Obviously, in polishing such an idea, we need to look at both sides of the duality equally well.
Not only do we need gravity dual computations, we also need to compute entanglement entropies directly within strongly-coupled CFTs.
Even so, the latter has been particularly lacking;
local observables are difficult to compute in strongly-coupled CFTs in $D>2$, let alone the entanglement entropy.

One of the few available methods to obtain physical quantities in strongly-coupled CFTs is the large-charge expansion \cite{Hellerman:2015nra,Alvarez-Gaume:2016vff,Monin:2016jmo}.
It works for generic strongly-coupled CFTs with a continuous global symmetry $G$, and allows for the computation of observables associated to operators at large global charge $Q\gg 1$ (or more generally of large representation).
To achieve this, the method uses an effective field theory (EFT) around a large vacuum expectation value (VEV), which is then used to compute desired observables \it via \rm the state-operator mapping.
Such an EFT, called the large-charge EFT, is organised in terms of the $1/Q$-expansion, and so we can extract physical quantities in a simple and controlled manner by leveraging $1/Q$ as a semi-classical expansion parameter.

In this paper, we compute information theoretic quantities in strongly-coupled CFTs using the large-charge expansion.
This will result in one of the first concrete, holography-free computations of quantum-information-theoretic quantities in strongly-coupled CFTs.
The quantity of our interest is the charged vacuum R\'enyi entanglement entropy on a disk, denoted as $S_n(Q)$.
It is defined for theories with a global symmetry (we will hereafter focus on $U(1)$) as the R\'enyi entropy of the density matrix $\tilde{\rho}_{A,Q}\equiv \rho_{A}\mathcal{P}_Q/\Tr\left[\rho_{A}\mathcal{P}_Q\right]$, where $\mathcal{P}_Q$ is a projection operator onto the charge-$Q$ subspace and $\rho_A$ is the reduced density matrix on region $A$ \cite{Belin:2013uta,Goldstein:2017bua}.
The quantity has interesting applications such as quantifying the speed of symmetry restoration after a symmetry-breaking quantum quench \cite{Feldman:2019upn,Oshima:2022yrw,scopa2022exact,Murciano:2023zvk,Caceffo:2024jbc}, where our result is potentially useful.

The large-charge expansion quite naturally fits in when computing $S_n(Q)$.
As we will explain in the main body of the text, $S_n(Q)$ can be computed from the charge-$Q$ thermal partition function on the hyperbolic disk spatial slice \cite{Casini:2011kv}.
We will see that the large-charge EFT can be used to compute it order-by-order in the $1/Q$-expansion.
This means that $S_n(Q)$ can be obtained for \emph{any} CFTs at large-charge, as long as we are able to write down the large-charge EFT.

Aside from being the first such computations, we will find several interesting and universal results.
For one thing, we find that the probability of realising charge-$Q$ vacuum fluctuations inside a disk scales as $e^{-Q^{3/2}}$ at large-$Q$, for any three-dimensional CFTs with a proper thermodynamic limit.
We also find that $S_n(Q)$ converges to a universal $O(1)$ number at large-$Q$, for \emph{any} CFTs described by the same large-charge EFT, called the \emph{large-charge universality class} hereafter \cite{Hellerman:2015nra}.
As a concrete example, such a number is computed for the large-charge universality class including the $O(2)$ Wilson-Fisher CFT.

The rest of the paper is organised as follows.
We start in Section \ref{sec:chargedRe'nyi} with a short account of the charged R\'enyi entropy, in particular its relation to the charged thermal partition function on a hyperbolic spatial slice.
We will also discuss at length the regularisation we use in this paper.
We will then give a brief review of the large-charge expansion in Section \ref{sec:largecharge}, using which we compute the charged R\'enyi entropy for a broad class of CFTs in Section \ref{sec:comp}.
We conclude in Section \ref{sec:outlook} with a summary of results and the outlook.

\section{Charged R\'enyi entropy}

\label{sec:chargedRe'nyi}

\subsection{Charge-resolved density matrix}

The charged R\'enyi entanglement entropy is a variant of the entanglement entropy for systems with a global symmetry, taken to be $U(1)$ hereafter for simplicity \cite{Belin:2013uta,Goldstein:2017bua,Kusuki:2023bsp}.
Let the reduced density matrix of a (uncharged) state restricted to a region $A$ be $\rho_A$.
In the presence of a symmetry, it can be decomposed into a direct sum of smaller matrices in each charged sector,
\begin{align}
    \rho_A=\bigoplus_{Q}\rho_{A,Q}.
\end{align}
Note that each $\rho_{A,Q}$ is \it not \rm a density matrix \it per se, \rm because $p(Q)\equiv \Tr\left[\rho_{A,Q}\right]\neq 1$. 
Rather, $p(Q)$ gives an overall probability of realising charge-$Q$ fluctuations in region $A$.
On the other hand, the charged $n$-th R\'enyi entanglement entropy $S_{n}(Q)$ is related to the normalised reduced density matrix proportional to $\rho_{A,Q}$, defined as
\begin{align}
    S_{n}(Q)\equiv \frac{1}{1-n}\log\left(\frac{\Tr\left[\rho_{A,Q}^n\right]}{\Tr\left[\rho_{A,Q}\right]^n}\right).
    \label{eq:def-chargedre'yi}
\end{align}

From now on, we will implicitly take the region $A$ as a disk of radius $R$ in $\mathbb{R}^{D-1}$, and the state in the total Hilbert space as vacuum, $\ket{0}$.
For later convenience we will take the radial coordinates $\mathbb{R}^{D-1}$ whose origin is at the centre of the disk, with the radial direction denoted as $r$.

\subsection{A conformal map to the hyperbolic disk}

As shown in \cite{Casini:2011kv}, the modular Hamiltonian on a disk of a CFT is given by the time-evolution operator of the same CFT on $H^{D-1}$. 
In other words, we have $\rho_A = {e^{-2\pi R H_{\mathrm{m}}}}/{\Tr\left[e^{-2\pi R H_{\mathrm{m}}}\right]}$,
where $H_{\mathrm{m}}$ is the Hamiltonian of the CFT defined on $H^{D-1}$. 
Likewise, $\rho_{A,Q}$ can be written as
\begin{align}
    \rho_{A,Q} = \frac{e^{-2\pi R H_{\mathrm{m}}}}{\Tr\left[e^{-2\pi R H_{\mathrm{m}}}\right]}\mathcal{P}_Q.
    \label{eq:charged-density-modular}
\end{align}
using the projection operator on the charge-$Q$ subspace, $\mathcal{P}_Q$.
By plugging this into \eqref{eq:def-chargedre'yi}, we see that the charged R\'enyi entropy is given by
\begin{align}
    S_{n}(Q)\equiv \frac{\log Z_n(Q)-n\log Z_1(Q)}{1-n},
    \label{eq:charged-re'nyi-modular}
\end{align}
where $Z_n(Q)\equiv \Tr\left[e^{-2\pi R n H_{\mathrm{m}}} \mathcal{P}_Q\right]$.
In other words, $S_n(Q)$ can be obtained from the charge-$Q$ partition function on a thermal geometry $S^1\times H^{D-1}$ with metric
\begin{align}
    ds^2&=d\tau^2+R^2\left(du^2+\sinh^2 u \,d\Omega_{D-2}^2\right),
    \label{eq:hyperbolic-frame}
\end{align}
at inverse temperature $2\pi R n$ and $2\pi R$.

\subsection{UV regularisation and the boundary condition}
\label{sec:bdy}

The thermal partition function on $H^{D-1}$ and hence $S_n(Q)$ are divergent as the spatial slice $H^{D-1}$ has infinite volume.
This is no pathology as it simply maps back to the UV divergence of the entanglement entropy in the original frame -- it just needs to be regularised.
But, in order for the resulting regularised reduced density matrix to have a proper interpretation as a collection of probabilities, it needs to have a discrete spectrum.
Therefore, not all regularisations are suited for our purpose.

The correct regularisation is as follows: modify the original state with a factorisation map that separates the subregion and its compliment physically with boundary conditions,
$\mathcal{M}_\delta:\,H\to H_A\otimes H_B$.\footnote{The factorisation scale $\delta$ is a physical scale which separates the subregion from its compliment, and can be thought of as different from the UV cutoff below which the theory is conformal. \emph{In particular, it makes complete sense to have modular energy as large as one wants, after fixing $\delta$.} For more about the definition of tensor factorisation of the CFT Hilbert space used to define the entanglement entropy and the importance of boundary conditions there, see \cite{Cardy:2010zs,Ohmori:2014eia,Hellerman:2021fla}.}
Let us, to simplify the argument, take $H_A$ and $H_B$ to be the CFT Hilbert spaces with conformal boundary conditions imposed on the surfaces defined by $r=R- \delta$ and $r=R\left(1-\delta/R\right)^{-1}$, respectively.\footnote{As we will see later, this corresponds to placing the boundary condition on a constant $u$ slice in the hyperbolic frame.}
We will defer specifying the boundary condition until after conformal transforming the geometry to $S^1\times H^{D-1}$.
By using such a map, the regularised reduced density matrix is defined as $\rho_{A,\delta}\equiv \Tr_{B}\left[M_\delta \rho M^\dagger_\delta\right]$, and this has discrete spectrum as promised for compact CFTs.

After all these steps, we can separate the regularised reduced density matrix into charged sectors as $\rho_{A,\delta}=\bigoplus_{Q}\rho_{A,Q}$.
This is important: we stress that it only makes sense to pick a map $M_\delta$ in advance and compare between different charged sectors of the same regularised reduced density matrix.
In other words, it only makes physical sense to fix $\delta/R$ to be finite (implicitly taken small as a UV cutoff, but still $O(Q^0)$ in particular) and to fix the boundary condition there, \emph{before} resolving the reduced density matrix into charged sectors.

Let us specify the boundary condition imposed by $\mathcal{M}_{\delta}$.
The conformal transformation maps the regularisation surface to the one near the boundary of $H^{D-1}$, defined by
\begin{align}
    u=\log\left(\frac{2R}{\delta}+1\right)\equiv u_0.
    \label{eq:boundaryhyperbolic}
\end{align}
In other words, it maps the Hilbert space $H_A$ to the one defined on a regularised hyperbolic disk $H^{D-1}_{\rm reg}$ with a boundary condition imposed at $u=u_0$.
Note that the boundary condition obviously needs to be $U(1)$-symmetric for the concept of charge to be well-defined in its presence.
We will take it to be Neumann for maximal simplicity in this paper.

The regularisation fixes the volume of the regularised hyperbolic space to be
$\mathop{{\tt Vol}}(H^{D-1}_{\rm reg})=\mathop{{\tt Vol}}(S^{D-2}_{\rm reg})\times \int_{0}^{u_0}du\,\sinh^{D-2}(u)$,
where $\mathop{{\tt Vol}}(S^{D-2}_{\rm reg})$ is the area of the $(D-2)$-sphere of radius $R$.
For later convenience we define a rescaled volume, ${V}_{D-1}\equiv \mathop{{\tt Vol}}(H^{D-1}_{\rm reg})/R^{D-1}$;
We will focus on $D=3$ hereafter, in which case we have 
\begin{align}
    V_2=2\pi \left(\frac{e^{u_0}}{2}-1+\frac{e^{-u_0}}{2}\right)
\end{align}

To sum up, we have reduced our problem into computing the charge-$Q$ thermal partition function $Z(\beta,Q)$ using the regularised modular Hamiltonian, $H_{\mathrm{m}}^{\mathrm{reg}}$, 
\begin{align}
    Z(\beta,Q)\equiv \Tr\left[e^{-\beta H_{\mathrm{m}}^{\mathrm{reg}}} \mathcal{P}_Q\right].
\end{align}
Our modular Hamiltonian is regulated with a Neumann boundary condition at $u=u_0$ on the hyperbolic disk, so it has a discrete spectrum.
We will, in the following sections, evaluate $Z(\beta,Q)$ in the large-$Q$ expansion.

\section{Large-charge expansion}
\label{sec:largecharge}

Let us now turn to the large-charge expansion \cite{Hellerman:2015nra}, with which we will compute $Z(\beta,Q)$ in the $1/Q$-expansion.
In order to illustrate the idea, let us for the moment pick the $O(2)$ Wilson-Fisher CFT in three dimensions as one of the simplest strongly-coupled CFTs with a $U(1)$ symmetry.

A crucial observation in the large-charge expansion is that physical quantities involving a large-charge ground state $\ket{Q}$ can be computed by using a saddle-point configuration which is helical in time, order-by-order in the $1/Q$-expansion.
Such a VEV will look like $\phi=Ae^{i\mu t}$ ($t$ is the Lorentzian time in \eqref{eq:hyperbolic-frame}, $t=-i\tau$) in the Wilson-Fisher case, and so the $O(2)$ symmetry (as well as the conformal symmetry) is spontaneously broken.
Note that this is consistent with the Neumann boundary condition we have imposed.\footnote{See \cite{Cuomo:2021cnb} for more about the effect of boundary conditions at large global charge.}
We are therefore entitled to write down an EFT using the Nambu-Goldstone (NG) boson $\chi$, where we have decomposed the field as $\phi=ae^{i\chi}$.

As discussed in \cite{Hellerman:2015nra}, the EFT can be written down order-by-order at large charge density, $\rho\gg R^{-2}$:
\begin{align}
    L=\tilde{c}_{3/2}\abs{\partial\chi}^3+\tilde{c}_{1/2}{\tt Ric}_{3}\abs{\partial\chi}+O(\rho^{-1/2}),
    \label{eq:3dconformalsuperfluid}
\end{align}
where $\abs{\partial\chi}\equiv \sqrt{-\partial_\mu\chi\partial^\mu\chi}$, ${\tt Ric}$ is the Ricci scalar of the underlying geometry (which is a constant $-(D-1)(D-2)/R^2$ for $\mathbb{R}\times H^{D-1}$), and $\tilde{c}_{3/2,\,1/2}$ are some Wilsonian coefficients.
Note that this action is to be evaluated around a linear background, $\chi=\mu t$.
The charge density, given as $\rho\equiv {\frac{Q}{\mathop{{\tt Vol}}(H^{2}_{\rm reg})}}$, is related to $\mu$ \it via \rm the Noether's theorem so that $\mu =O\left(\sqrt{\rho}\right)$.
The subscripts of $c_*$ therefore denote the scaling at which the terms contribute, \it e.g., \rm $\tilde{c}_{3/2}$ is the Wilsonian coefficient at $O(\rho^{3/2})$.

Leaving coefficients such as $\tilde{c}_{3/2,\,1/2}$ undetermined is an unfortunate feature of the effective action, but it has its own benefits.
Take any theory with the $O(2)$ symmetry; if we know that the only light mode around the large-charge saddle-point is the NG boson, it is then immediate, because of the underlying conformal and $O(2)$ symmetries, that the \emph{same} effective action (with different Wilsonian coefficients) describes the large-charge sector of that theory.
We can express this by saying that theories with the same large-charge EFT define a \emph{large-charge universality class}.
For example, \eqref{eq:3dconformalsuperfluid} describes not only the $O(2)$ Wilson-Fisher CFT at large charge, but the $O(N)$ Wilson-Fisher CFT at large-rank symmetric traceless representations \cite{Alvarez-Gaume:2019biu} and $\mathbb{C}P^{N-1}$ model at large monopole number \cite{DeLaFuente:2018uee}.

\section{Large-charge R\'enyi entropy}

\label{sec:comp}



\subsection{Dispersion relation}

Let us start by deriving the dispersion relation on top of our large-charge vacuum, $\chi=\mu t$.
\red{This will serve as a starting point for computing the thermal partition function $Z(\beta,Q)$, as the resulting one-particle spectrum governs the leading contributions to the energy spectrum in the large-$Q$ limit (see \eqref{eq:oneloop}, for example.)}
By separating $\chi$ into VEV and fluctuation, $\chi=\mu t+\hat{\chi}$, the fluctuation Lagrangian at leading order, which is $O(1)$, becomes
\begin{align}
    L\ni\hat\chi\left(-\partial_{t}^2+\frac{1}{2R^2}\triangle_{H^{2}_{\rm reg}}\right)\hat{\chi},
    \label{eq:fluctuation-lagrangian}
\end{align}
where $\triangle_{H^2_{\rm reg}}\equiv \frac{1}{\sinh u}\partial_u\left(\sinh u\partial_u\right)+\frac{1}{\sinh^2 u}\partial_\theta^2$ is the hyperbolic Laplacian, with the subscript $\rm reg$ indicating $\hat{\chi}$ needs to satisfy the Neumann boundary condition.

The leading order dispersion relation can be derived by solving the equation of motion (EOM) of the quadratic fluctuation Lagrangian:
\begin{align}
    \omega_{m,k}^2R^2=\frac{1}{2}\left(\frac{1}{4}+s^2_{m,k}\right)
    \label{eq:111}
\end{align}
where $\frac{1}{4}+s^2_{m,k}$ denotes the $k$-th eigenvalue of $\triangle_{H^2_{\rm reg}}$ in the sector of angular momentum $m$, and the corresponding eigenfunction is $\chi_{m,k}(u,\theta)\equiv P^{m}_{-1/2+is_{m,k}}(\cosh u)e^{im\theta+i\pi}\in \mathbb{R}$.
As already implicit, this one-particle spectrum is discrete instead of continuous because of the Neumann boundary condition.\footnote{One might think that the continuous spectrum suitably approximates the discrete spectrum, but that would be too quick. Because the IR cutoff is always kept finite, at large enough angular momentum $m$, the continuous description is never correct. In fact we would get a wrong coefficients for $A_n$ had we used the continuous approximation.}
\red{It is extremely important that the spectrum, at leading order in large-$Q$, does \emph{not} depend on the Wilsonian coefficients, $c_{*}$, \emph{at all}.
This fact will translate to the universality of the charged R\`enyi entropy at large-$Q$, as shown in \eqref{eq:final}.}

Because the discrete spectrum is hard to obtain analytically, we have computed $\omega_{m,k}$ numerically for various $e^{u_0} \gg 1$.\footnote{\red{They were obtained by locating the zeros of $\partial_u \chi_{m,k}(u,\theta)$ by using a bisection method. See \url{https://github.com/machiparu/lcre} for the Python program.}}
This is in anticipation that the final result will be organised in terms of $e^{u_0}$, which will be justified \it a posteriori \rm when we compute the R\'enyi entropy.
We also truncated the computation at $s_{m,k}=3$ because each mode labelled by $s_{m,k}\gtrsim 3$ will contribute to the R\'enyi entropy at order $O(e^{-2\pi \omega_{m,k}})=O(10^{-6})$.
We show an example of the spectrum for $u_0=3$ in Table \ref{tab:spectrum}.

\begin{table*}[t]
\centering
\setlength{\extrarowheight}{0pt}
\addtolength{\extrarowheight}{\aboverulesep}
\addtolength{\extrarowheight}{\belowrulesep}
\setlength{\aboverulesep}{0pt}
\setlength{\belowrulesep}{0pt}
\caption{List of $s_{m,k}\leq 3$, which gives the one-particle spectrum subject to the Neumann boundary condition \emph{via} \eqref{eq:111}, at $u_0=3$.}
\label{tab:spectrum}
\resizebox{\textwidth}{!}{%
\begin{tabular}{ccccccccccccccccccccccc} 
\toprule
          & $m=0$  & $\abs{m}=1$  & $\abs{m}=2$  & $\abs{m}=3$  & $\abs{m}=4$  & $\abs{m}=5$  & $\abs{m}=6$  & $\abs{m}=7$  & $\abs{m}=8$  & $\abs{m}=9$  & $\abs{m}=10$ & $\abs{m}=11$ & $\abs{m}=12$ & $\abs{m}=13$ & $\abs{m}=14$ & $\abs{m}=15$ & $\abs{m}=16$ & $\abs{m}=17$ & $\abs{m}=18$ & $\abs{m}=19$ & $\abs{m}=20$ & $\abs{m}\geq 21$                      \\ 
\midrule
$k=1$     & $1.21$ & $1.66$ & $0.26$ & $0.54$ & $0.74$ & $0.91$ & $1.07$ & $1.22$ & $1.36$ & $1.50$ & $1.63$ & $1.77$ & $1.90$ & $2.02$ & $2.15$ & $2.27$ & $2.40$ & $2.52$ & $2.64$ & $2.76$ & $2.88$ & {\cellcolor{grayrow}}$s_{m,k}>3$  \\
$k=2$     & $2.30$ & $2.78$ & $2.00$ & $2.28$ & $2.53$ & $2.76$ & $2.98$ & \multicolumn{15}{c}{{\cellcolor{grayrow}}$s_{m,k}>3$}                                                                                                           \\
$k\geq 3$ & \multicolumn{22}{c}{{\cellcolor{grayrow}}$s_{m,k}>3$}                                                                                                                                                                          \\
\bottomrule
\end{tabular}
}
\end{table*}

\subsection{Spectra at large-charge}

Let us now use these data to compute the ground state energy $E_{\rm g.s.}(Q)$ at large-$Q$,
which is simply given by a sum of vacuum diagrams around the VEV.
This is a controlled expansion as the action evaluated on the VEV is $O(\left({Q}/{V_2}\right)^{3/2})$, which works as the loop-counting parameter at large-$Q$.
Up to $O(1)$ and above, the result consists of tree-level and one-loop, resulting in
\begin{align}
	{{E_{\rm g.s.}(Q)}R}={V_2}\left(c_{3/2}\,q^{\frac{3}{2}}+c_{1/2}\,q^{\frac{1}{2}}+\alpha+O(q^{-\frac{1}{2}})\right),
\end{align}
where $q\equiv Q/V_2$. 
Note that we are taking $Q\to \infty$ limit first, while taking $V_2$ to be large but fixed in terms of $Q$ (see Section \ref{sec:bdy}).
While $c_{*}$ are the Wilsonian coefficients and cannot be determined from the EFT analysis alone, the one-loop contribution $\alpha$ is a number that is in principle calculable.\footnote{\red{For the $O(2)$ Wilson-Fisher CFT, one can use the operator dimensions of $\phi^Q$ obtained \emph{via} the Monte-Carlo simulation \cite{Loukas:2017lof,Banerjee:2019jpw,Cuomo:2023mxg} to read out $(\tilde{c}_{3/2},\tilde{c}_{1/2})=(0.103,-0.020)$. This translates to $({c}_{3/2},{c}_{1/2})=(1.20,-0.370)$.}}
While it is universal among theories in the same large-charge universality class and hence is in general important \cite{Hellerman:2015nra,Cuomo:2020rgt}, its precise value is of no use to us later on -- we leave its computation for future work.

The excitation spectrum can be inferred easily from the one-particle spectrum.
It is labelled by the number of excited phonon modes $N_{m,k}$ (or collectively $\vec{N}$) with energy $\omega_{m,k}$ as $\Delta E_{\vec{N}}=\sum_{m,k} N_{m,k}\omega_{m,k}$.
The EFT cutoff puts a bound on $\Delta E_{\vec{N}}$ at order $O(\sqrt{q})$, above which such a computation should never be trusted.
Even then, because what we are going to compute is the thermal partition function, we will only need $O(1)$ excitation energy (anything scaling with $Q$ will get exponentially damped in the limit we are considering), where our EFT analysis is perfectly valid at leading order.

\subsection{Large-charge fluctuation probability}

Let us compute $p(Q)\equiv \Tr\left[\rho_{A,Q}\right]$ which gives the total probability of measuring charge-$Q$ fluctuations on a disk of radius $R$ in vacuum.
We can use \eqref{eq:charged-density-modular} to rewrite $p(Q)$ as
\begin{align}
    p(Q)=\frac{\Tr\left[e^{-2\pi R H_{\mathrm{m}}}\mathcal{P}_Q\right]}{\Tr\left[e^{-2\pi R H_{\mathrm{m}}}\right]}
    =\frac{Z(2\pi R,Q)}{Z(2\pi R)},
\end{align}
where $Z(\beta)$ is the full thermal partition function at inverse temperature $\beta$, \it i.e., \rm $Z(\beta)\equiv \sum_{Q}Z(\beta,Q)$.

The numerator can be computed by using the knowledge of the one-particle spectrum $\omega_{m,k}$
because the charged thermal partition function at temperature $1/\beta$ is given by
\begin{align}
    \log Z(\beta, Q)=-\beta E_{\rm g.s.}(Q)-\sum_{m,k}\log\left(1-e^{-\beta \omega_{m,k}}\right).
    \label{eq:oneloop}
\end{align}
On the other hand, the denominator $Z(2\pi R)$ is some $O(1)$ number which is difficult to compute for generic strongly-coupled theories.
In computing $\log p(Q)$, therefore, up to and above $O(1)$, we have
\begin{align}
    \log p(Q)=-2\pi {V_2}\left(c_{3/2}\,q^{\frac{3}{2}}+c_{1/2}\,q^{\frac{1}{2}}+c_0+O(q^{-\frac{1}{2}})\right).
\end{align}
Unlike the case of the ground state energy, the $O(1)$ contribution $c_0$ is theory-dependent, consisting of the one-loop Casimir energy $\alpha$ and the one-loop thermal contribution which are universal, and $\log Z(\beta)$ which is theory-dependent.

Note that the result is intuitive in that the probability of measuring charge $Q$ inside a ball-shaped region in vacuum is exponentially damped at large-$Q$.
Furthermore, the relation $p(Q)=O(e^{-q^{3/2}})$ is presumably universal for any three-dimensional theories with a proper thermodynamic limit because of dimensional analysis \cite{Hellerman:2015nra}.
This is true even when they are not described by \eqref{eq:3dconformalsuperfluid} at large-charge, \it e.g., \rm the free Fermion theory \cite{Komargodski:2021zzy}.

\subsection{Large-charge entanglement entropy}

The charged R\'enyi entropy can be written using the charged thermal partition function, as shown in \eqref{eq:charged-re'nyi-modular},
\begin{align}
    S_{n}(Q)\equiv \frac{\log Z(2\pi Rn, Q)-n\log Z(2\pi R,Q)}{1-n}.
    \label{eq:charged-re'nyi-modular2}
\end{align}
Thus it is immediate that the vacuum contribution in \eqref{eq:oneloop} cancels out, and that it depends only on the one-particle spectrum, $\omega_{m,k}$.
It is also apparent by now that $S_n(Q)$ has an expansion in terms of $1/q$, and that its leading order is $O(1)$, which
simply comes from the fluctuation Lagrangian \eqref{eq:fluctuation-lagrangian}.
Interestingly, this number is universal across different CFTs as long as they are in the large-charge universality class described by \eqref{eq:3dconformalsuperfluid};
The only theory dependence lies in subleading corrections.

\begin{figure}[t!]
    \centering
    \caption{A plot of the large-charge von Neumann entropy $S_{n=1}^{(0)}(Q)$ in terms of $u_0$ (red dots) with a comparison to a fit (blue line). We see a perfect fit at large $e^{u_0}$. 
    }
    \includegraphics[width=0.9\linewidth]{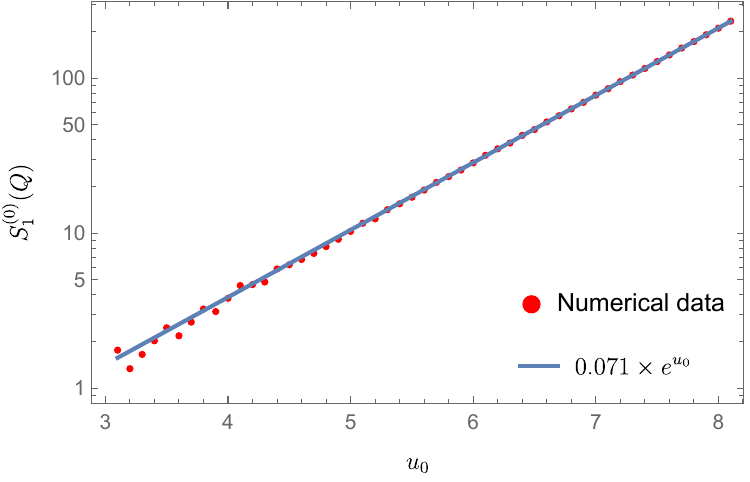}
    \label{fig:vNplotu_0}
\end{figure}

We now numerically compute the leading $O(1)$ part of $S_n(Q)$, denoted as $S_n^{(0)}(Q)$, using the numerically obtained values of $\omega_{m,k}$.
As an example, we show a plot of $S_{n=1}^{(0)}(Q)$ as a function of $u_0$ in Figure \ref{fig:vNplotu_0}; we see that it behaves as $O(e^{u_0})$ at large $u_0$, which continues to be true for other values of $n$.\footnote{This is by no means trivial. With a physical boundary condition imposed on the hyperbolic space, there is no guarantee that the R\'enyi entropy scales with the regularised volume of the hyperbolic space, as is usually claimed.}



\begin{figure}[t!]
    \centering
    \caption{A plot of $A_n$ in terms of $n$ (red dots) with a comparison to a fit (blue line). 
    }
    \includegraphics[width=0.9\linewidth]{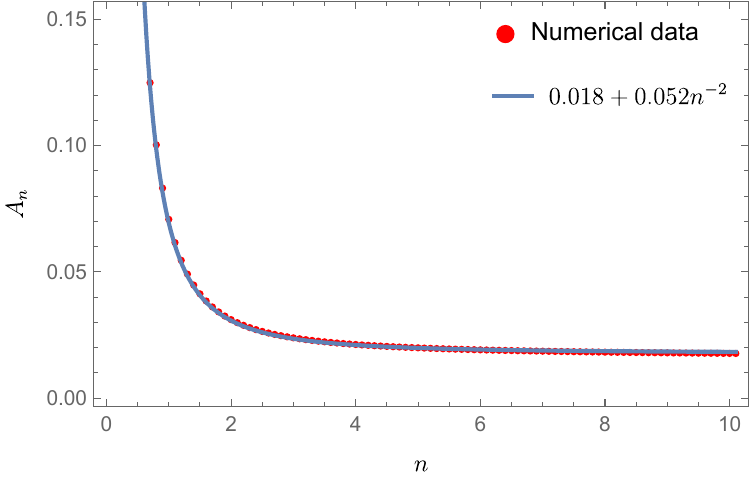}
    \label{fig:re'nyifinalplot}
\end{figure}

By fitting the numerical value of $S_n^{(0)}(Q)$ against an expansion in terms of $V_2=2\pi \left(\frac{e^{u_0}}{2}-1+\frac{e^{-u_0}}{2}\right)=2\pi\frac{R}{\delta}+O(1)$, we see that\footnote{Note that this is not the usual expansion where $V_2$ is the largest parameter. in which the $O(1)$ part is commonly argued to be universal.}
\begin{align}
    S_n(Q)=\left(A_{n}\pi^{-1}V_2+O({V_2}^{0})\right) q^{0}+O(q^{-1/2}),
    \label{eq:final}
\end{align}
where $A_{n}$ is a numerical function plotted in Figure \ref{fig:re'nyifinalplot};
For example, we find
\begin{align}
    A_1=0.071,\quad A_2 = 0.031, \quad A_3 = 0.024.
\end{align}
To reiterate, these numbers are universal for any CFTs in our large-charge universality class.

Incidentally, $A_n$ is approximated well by $A_n\approx 0.052n^{-2}+0.018$ at small $0.5\lesssim n \lesssim 5$.
To see why, notice that \eqref{eq:fluctuation-lagrangian} is that of a minimally coupled free scalar, and so its 
high-temperature thermal partition function can be approximated by that of a conformal scalar with corrections that go as $O(\beta)$. 
This allows us to use a general analysis of high-temperature limit of CFTs using thermal EFTs \cite{Benjamin:2023qsc,Kusuki:2025pgx}, which leads to $S_n=f/n^2+g/n+\text{(const.)}$, consistent with our fit.
Note that $g$ is related to the boundary entropy and so its absence could be explained by the use of the Neumann boundary condition.\footnote{We thank Sridip Pal for discussions on this paragraph.}

\section{Conclusions and Outlook}

\label{sec:outlook}

In this paper, we computed the charged (symmetry-resolved) vacuum R\'enyi entanglement entropy on a disk in a certain class of strongly-coupled CFTs in three dimensions, by taking the limit of large global charge.
The starting point of the computation was the conformal transformation, which relates the charged R\'enyi entropy to a thermal partition function on a (regulated) hyperbolic space.
Then the large-charge EFT was introduced, which was argued to completely control the $O(1)$ spectrum above the large-charge vacuum, as a tool to compute the thermal partition function.
We then proceeded to compute the large-charge R\'enyi entropy, taking an example of the $O(2)$ Wilson-Fisher CFT; We found that it converges to an $O(1)$ number at large charge, and moreover that it is universal across any CFTs described by the same large-charge EFT, including \it e.g., \rm the $O(N)$ Wilson-Fisher CFT at large-rank symmetric traceless representations and $\mathbb{C}P^{N-1}$ model at large monopole number.

There are a number of interesting future directions.
First of all, it would be interesting to compute the large-charge R\'enyi entropy for various other theories.
For example, SUSY theories with a moduli space are described by free complex scalars at leading order in large-charge \cite{Hellerman:2017veg,Hellerman:2018xpi,Hellerman:2020sqj,Hellerman:2021yqz,Hellerman:2021duh,Heckman:2024erd}, so we expect a completely different spectrum of the modular Hamiltonian at large charge than the present case.
It would also be interesting to compute it using perturbative parameters (large-$N$, $\epsilon$-expansion, etc.) in the double-scaling limit (fixing $Q/N$, $\epsilon Q$, for example) \cite{Alvarez-Gaume:2019biu,Bourget:2018obm,Beccaria:2018xxl,Beccaria:2018owt,Hellerman:2018xpi,Grassi:2019txd,Watanabe:2019pdh,Arias-Tamargo:2019xld,Badel:2019oxl,Sharon:2020mjs,Watanabe:2022htq,Beccaria:2020azj,Antipin:2020abu,Antipin:2020rdw}.
Additionally, even though none of the theories we discussed in this paper is holographic, it would be interesting to see to what extent our current result can be reproduced using holography \cite{Belin:2013uta,Loukas:2018zjh,delaFuente:2020yua,Choi:2025tql}.

We can also study various other limits of the symmetry-resolved entanglement entropy, or other interesting symmetry-enriched entanglement quantities such as entanglement asymmetry \cite{Fossati:2024xtn,Fossati:2024ekt,Kusuki:2024gss}.
For example, we can take the large-representation limit in a theory with a non-Abelian symmetry.
This would result in various one-particle excitations containing non-relativistic NG bosons as well as phonons \cite{Alvarez-Gaume:2016vff,Hellerman:2017efx,Hellerman:2018sjf,Watanabe:2019adh,Cuomo:2021qws}.
Another direction would be to study the limit of large spin, which is known to be an interesting solvable limit for operator dimensions and the OPE coefficients \cite{Fitzpatrick:2012yx,Komargodski:2012ek,Alday:2016njk,Alday:2016jfr,Dey:2017fab,Pal:2022vqc,vanRees:2024xkb}.
It would be interesting to ask if it also translates to the simplification of the modular Hamiltonian spectrum.
One can also study the regime of large-charge \emph{and} large-spin, which is known to exhibit various interesting phases with \it e.g., \rm vortex lattices and the giant vortex \cite{Cuomo:2017vzg,Kravec:2019djc,Cuomo:2019ejv,Cuomo:2022kio,Choi:2025tql}.

\section*{Acknowledgements}

The author thanks Simeon Hellerman, Kantaro Ohmori, Sridip Pal, and Adar Sharon
for valuable discussions.
This work is supported by a Grant-in-Aid for JSPS Fellows No.~22KJ1777, a Grant-in-Aid for Early-Career Scientists No.~25K17387, and by a MEXT KAKENHI Grant No.~24H00957.

\appendix

\bibliographystyle{JHEP}
\bibliography{main,ref}

\end{document}